\documentclass{jaa}
\usepackage[colorlinks=true,citecolor=blue]{hyperref}
\usepackage[authoryear,round]{natbib}

\usepackage{graphicx}
\usepackage{parskip}

\begin{document}\sloppy

\title{The statistical analysis of the dynamical evolution of the open clusters}


\author{Jayanand Maurya\textsuperscript{1,*}, Y. C. Joshi\textsuperscript{2}, Manash Ranjan Samal\textsuperscript{1}, Vineet Rawat\textsuperscript{1}, and Anubha Singh Gour\textsuperscript{3}}
\affilOne{\textsuperscript{1}Astronomy \& Astrophysics Division, Physical Research Laboratory, Ahmedabad, 380009, State of Gujarat, India\\}
\affilTwo{\textsuperscript{2}Aryabhatta Research Institute of observational sciencES (ARIES), Nainital, Uttrakhand, India\\}
\affilThree{\textsuperscript{3}School of Studies in Physics and Astrophysics, Pandit Ravishankar Shukla University, Chattisgarh 492 010, India}


\twocolumn[{

\maketitle

\corres{maurya.jayanand@gmail.com}


\begin{abstract}We present the dynamical evolution of ten open clusters which were part of our previous studies. These clusters include both young and intermediate-age open clusters with ages ranging from 25 $\pm$ 19 Myr to 1.78$\pm$0.20 Gyr. The total mass of these clusters ranges from 356.18$\pm$142.90 to 1811.75$\pm$901.03 M$_{\odot}$. The Galactocentric distances to the clusters are in the range of  8.91$\pm$0.02 to 11.74$\pm$0.18 kpc.  The study is based on the ground-based UBVRI data supplemented by the astrometric data from the Gaia archive. We studied the minimum spanning tree of the member stars for these clusters. The mass segregation in these clusters was quantified by mass segregation ratios calculated from the mean edge length obtained through the minimum spanning tree. The clusters NGC 2360, NGC 1960, IC 1442, King 21, and SAI 35 have $\Gamma_{MSR}$ to be 1.65$\pm$0.18, 1.94$\pm$0.22, 2.21$\pm$0.20, 1.84$\pm$0.23, and 1.96$\pm$0.25, respectively which indicate moderate mass segregation in these clusters. The remaining five clusters are found to exhibit weak or no mass segregation. We used the ratio of half mass radius to the tidal radius i.e. R$_{h}$/R$_{t}$ to investigate the effect of the tidal interactions on the cluster structure and dynamics. The ratios of half mass radii to tidal radii are found to be positively correlated with the Galactocentric distances with a linear slope of 0.06$\pm$0.01 having linear regression coefficient r-square = 0.93 for the clusters.
\end{abstract}

}]


\doinum{12.3456/s78910-011-012-3}
\artcitid{\#\#\#\#}
\volnum{000}
\year{0000}
\pgrange{1--}
\setcounter{page}{1}
\lp{1}

\section{Introduction}
The star clusters evolve dynamically as the age passes after their birth from molecular clouds. Open clusters go through two-body relaxation and their core approaches spherical structure due to the dynamical relaxation. This dynamical relaxation causes low-mass stars to preferentially settle in the outer part of the clusters. These outskirt stars are subjected to dynamic ejection from the cluster which lowers the gravitational potential and loosens the cluster \citep{2008MNRAS.389L..28G,2019ApJ...877...12T}. The external perturbations like tidal effects, disc crossing, and differential rotation also strengthen the disintegration process of the open clusters. These internal and external factors affect the dynamical evolution of the clusters which manifests in the morphology, shape, and spatial distribution of the stars of the clusters. Thus, open clusters with accurately determined physical parameters are useful in understanding the impact of the internal and external effects on the dynamical evolution of the cluster.

The spatial distribution of stars in the clusters has been found to be dependent on the stellar masses. The massive stars are found to be preferentially concentrated in the inner region of the clusters compared to low-mass stars \citep{2018MNRAS.473..849D,2020MNRAS.499..618J,2021AJ....162...64M}. This mass segregation phenomenon has been explained by two theories that attribute mass segregation either to the star formation process itself or dynamical evolution. The theory considering the star formation process responsible for the mass segregation proposes that massive stars are preferentially formed in the inner region of the cluster \citep{2008ApJ...678L.105D}. However, the dynamical evolution theory of mass segregation suggests that the segregation happens due to an internal two-body dynamical relaxation process \cite{2009ApJ...700L..99A}. The debate on the origin of mass segregation has not concluded and requires further analysis \citep{2018MNRAS.473..849D}.

The comprehensive study of the cluster parameters like core radii, tidal radii, half-mass radii, and ages helps in understanding the complete scenario of the interplay among structural evolution, dynamical evolution, and tidal interactions. Open clusters generally become more centrally concentrated due to dynamical evolution which makes mass loss caused by tidal interactions in these clusters less likely. In this two-body dynamical relaxation, massive stars sink towards the central region while low-mass stars preferentially shift towards the outer region which evaporates from the cluster with passing time. 

The impact of the tidal field on the cluster size can be characterized by using the ratio of half-mass radius to the tidal radius of the cluster i.e. R$_{h}$/R$_{t}$. Previously, it has been found that the R$_{h}$/R$_{t}$  ratio parameter is also related to the survival of the cluster as it has been found that the more compact clusters have smaller R$_{h}$/R$_{t}$ values which favour the survival of the clusters located at smaller Galactocentric distances by reducing the  mass loss due to tidal fields \citet{2021MNRAS.500.4338A}.

To probe the role of dynamical evolution in defining the shape and morphology of the clusters, we present the analysis of the dynamical evolution of the ten open clusters. These clusters are parts of our previous studies \citep{2020MNRAS.499..618J,2020MNRAS.494.4713M,2020MNRAS.495.2496M,2021AJ....162...64M} and their physical parameters such as reddening, age, and distance are listed in the Table~\ref{clust_par}. In the current study, we applied a statistical approach to study the dynamical evolution of this homogeneous sample of open clusters.  
\section{Data}

The observations for the clusters were taken using the 1.3-m Devasthal Fast Optical Telescope (DFOT) at Devasthal and 1.04-m Sampurnanand Telescope (ST) at Nainital. The telescopes were equipped with 2k$\times$2k CCDs having a field of view of $\sim$ $18^{\prime}$ $\times$ $18^{\prime}$ and $\sim$ $13^{\prime}$ $\times$ $13^{\prime}$ for the DFOT and ST telescopes, respectively. The plate scales were 0.54 and 0.75 arcsec pixel$^{-1}$ for the DFOT and ST telescopes, respectively. The observations were taken on the nights of 30 November 2010, 24 \& 25 March 2017, 21 October 2017, and 13 January 2018.   We used the Image Reduction and Analysis Facility (IRAF) for cleaning the photometric data. We used the PSF technique to obtain the instrumental magnitudes using DAOPHOT $\MakeUppercase{\romannumeral 2}$ software packages. The instrumental magnitudes were calibrated to standard magnitudes through the process described by \citet{1992ASPC...25..297S}. The conversion formula used for the standardization of instrumental magnitudes is given in our previous paper \citet{2020MNRAS.494.4713M}. The membership of these clusters was calculated using proper motions and parallaxes from Gaia DR3 archives. We first plotted the vector-point diagram (VPD) of these clusters using proper motions. We noticed over-density regions in the VPDs of the clusters which were found to be formed by probable cluster members on the basis of the location of the stars belonging to these regions on the Hertzsprung–Russell (H-R) diagrams. To quantify the membership probabilities of stars of the clusters, we calculated the membership probabilities from the proper motions of the stars using a statistical approach originally suggested by \citet{1971A&A....14..226S}. The membership probability for the i$^{th}$ star was calculated as follows:\\
$$P (i) = \frac{n_{c}~.~\phi_c^{\nu}(i)}{n_{c}~.~\phi_c^{\nu}(i) + n_f~.~\phi_f^{\nu}(i)}$$
, where $n_{c}$ and $n_{f}$ denote the number of stars belonging to field and cluster regions in the normalized form such that $n_{c}$ + $n_{f}$ = 1. The $\phi_c^{\nu}(i)$ and $\phi_f^{\nu}(i)$ in the above equation are the frequency distribution functions for the stars belonging to cluster and field populations, respectively. We calculated the $\phi_c^{\nu}(i)$ and $\phi_f^{\nu}(i)$ from the $i^{th}$ stars' proper motions($\mu_{\alpha*i}$; $\mu_{\delta i}$), errors in the proper motions ($\epsilon_{\alpha* i}$; $\epsilon_{\delta i}$), and the mean proper motion of probable cluster members or field stars ($\mu_{\alpha* c}$; $\mu_{\delta c}$) with their dispersions ($\sigma_{\alpha* c}$; $\sigma_{\delta c}$) as described in \citet{2020MNRAS.494.4713M}. We also used Gaia parallaxes to remove the possible contamination of member stars from field stars with similar proper motions. The number of stars, N, identified as member stars up to the completeness limit of the data for each cluster is given in Table~\ref{clust_par}. The ADDSTAR routine of DAOPHOT packages was utilized to estimate the completeness of the data by adding artificial stars in the original images of the clusters. The completeness of the data was measured based on the ratio of the number of recovered stars to the number of artificially added stars in each magnitude bin. The completeness determination process is described in our previous study \citep{2020MNRAS.494.4713M}. The completeness limit of the data in the V band is denoted by V$_{lim}$ in Table~\ref{clust_par}. 

The physical parameters of the studied clusters are derived through the samples of member stars identified by us. The photometric data using our observations are complemented by Gaia DR3 and Pan-STARSS data for this study. The physical parameters derived in our previous studies such as reddening, extinction law, distance, ages, and completeness limits were used for the present study to be uniform in our sample as well as our approach  \citep{2020MNRAS.499..618J,2020MNRAS.494.4713M,2020MNRAS.495.2496M,2021AJ....162...64M}.\\

The reddening of all the clusters except SAI 35, SAI 44, and SAI 45 was estimated through fitting zero-age main sequence isochrones given by \citet{Schmidt-Kaler} on the (U-B)/(B-V) colour-colour diagrams. In the absence of the U band data, we calculated the reddening of the clusters SAI 35, SAI 44, and SAI 45 using a 3D reddening map given by \citet{2019ApJ...887...93G}.  We calculated E(B-V)  values for the clusters from the E(g-r) values provided by the reddening map using the extinction ratios relations of \citet{2019ApJ...877..116W}. The clusters included in this study are associated with relatively low reddening for the detection of lower mass regime member stars. The total-to-selective extinction values for the clusters were determined using (V-$\lambda$)/(B-V) two-colour diagrams where $\lambda$ = R, I, J, H, K bands magnitudes. The near-IR bands' magnitudes were obtained from 2MASS archives. The distances to the clusters were estimated using \textit{Gaia} DR3 parallaxes through the parallax inversion method as described in our previous study \citep{2020MNRAS.495.2496M}. The distances to the clusters are calculated to be in the range of 1072$\pm$44 to 3670$\pm$184 parsecs. The ages of the clusters were estimated by fitting \citet{2017ApJ...835...77M} isochrones on the colour-magnitude diagrams for derived reddening and distance modulus. The ages of the clusters were estimated in the range of 25 $\pm$ 19 Myr to 1.78$\pm$0.20 Gyr. The methods for the derivation of the physical parameters are briefly described in \citet{2020MNRAS.495.2496M,2021AJ....162...64M} and their values are given in Table~\ref{clust_par}.\\
\begin{table*}\fontsize{10}{10}\selectfont
  \centering
  \caption{The derived values of physical parameters of the clusters. The name, right ascension, declination, completeness limit in the V band, the mass of the most massive star, the mass completeness limit of the data, number of member stars up to the completeness limit of the data, reddening, logarithmic age, and distance calculated using the methods described by \citet{2018AJ....156...58B} are given in columns 1 to 10, respectively.}
  \label{clust_par}
  \begin{tabular}{c c c c c c c c c c}  
  \hline  
    Cluster& RA& Dec& V$_{lim}$& M$_{upper}$& M$_{lower}$& N  & E(B-V) &  log(Age) & D$_{BJ}$ \\
           &   &   &  (mag)   &  M$_{\odot}$& M$_{\odot}$&    & (mag)    &  (Myr)    & (pc) \\
  \hline
  NGC 381&  01:08:19.6& +61:35:18.2& 20 &  2.80& 0.61& 134&0.36$\pm$0.04& 8.65$\pm$0.05& 1147$\pm$38 \\
  NGC 2360& 07:17:43.5& -15:38:39.8& 18 &  2.21& 0.68& 276&0.07$\pm$0.03& 8.95$\pm$0.05& 1072$\pm$44 \\
  Berkeley 68&    04:44:30.0& +42:05:55.8&  19 & 1.75& 0.97& 229&0.52$\pm$0.04& 9.25$\pm$0.05& 3206$\pm$199\\ 
  NGC 1960& 05:36:20.2&  +34:08:06& 19 & 7.18& 0.72& 253&0.24$\pm$0.02& 7.44$\pm$0.02& 1169$\pm$54 \\
  IC 1442&  22:16:03.7& +53:59:29.4& 18 & 9.02& 1.43& 205&0.54$\pm$0.04&7.40$\pm$0.30& 3492$\pm$230 \\
  King 21&  23:49:55.0& +62:42:18.0& 19 & 6.89& 1.21& 238&0.76$\pm$0.06&7.70$\pm$0.20& 2953$\pm$174 \\
  Trumpler 7&     07:27:23.8& -23:56:56.4& 19 & 5.12& 0.76& 146&0.38$\pm$ 0.04&7.85$\pm$0.25& 1700$\pm$98 \\
  SAI 35&   04:10:46.8& +46:52:33.2& 19 & 3.34& 1.02& 156&0.61$\pm$ 0.04&8.50$\pm$0.10& 2826$\pm$266 \\
  SAI 44&   05:11:10.5& +45:42:10.2& 19 & 2.47& 1.00& 171&0.34$\pm$ 0.04&8.82$\pm$0.10& 3670$\pm$184 \\
  SAI 45&   05:16:29.4& +45:35:35.9& 19 & 2.14& 0.79&  79&0.34$\pm$ 0.02&9.07$\pm$0.10& 1668$\pm$47 \\
  \hline
  \end{tabular}
\end{table*}
\section{Results}
\subsection{Mass segregation}

The mass segregation can be attributed to the escape of the low-mass stars from the cluster besides the concentration of massive stars in the central region of the cluster. The mass segregation effect can be triggered due to the dynamical evolution through the equipartition of energy or maybe imprint of the star formation process itself \citep{1988MNRAS.234..831S}. The cumulative distribution of stars with a radius for various mass ranges is often used to study the mass segregation in star clusters. To study the effect of dynamical evolution and mass segregation, we determined cumulative radial distributions of member stars for different mass ranges in our previous studies \citep{2020MNRAS.499..618J,2020MNRAS.494.4713M,2020MNRAS.495.2496M}. However, the method based on the cumulative distribution of stars with radial distance depends on the size of mass bins and cumulative radii which may give misleading results. Therefore, we used a method given by \citet{2009MNRAS.395.1449A} which is based on mass segregation ratio (MSR) for the study of mass segregation scenario in the clusters.

\begin{figure*}
\centering
\includegraphics[width=19 cm, height=9 cm]{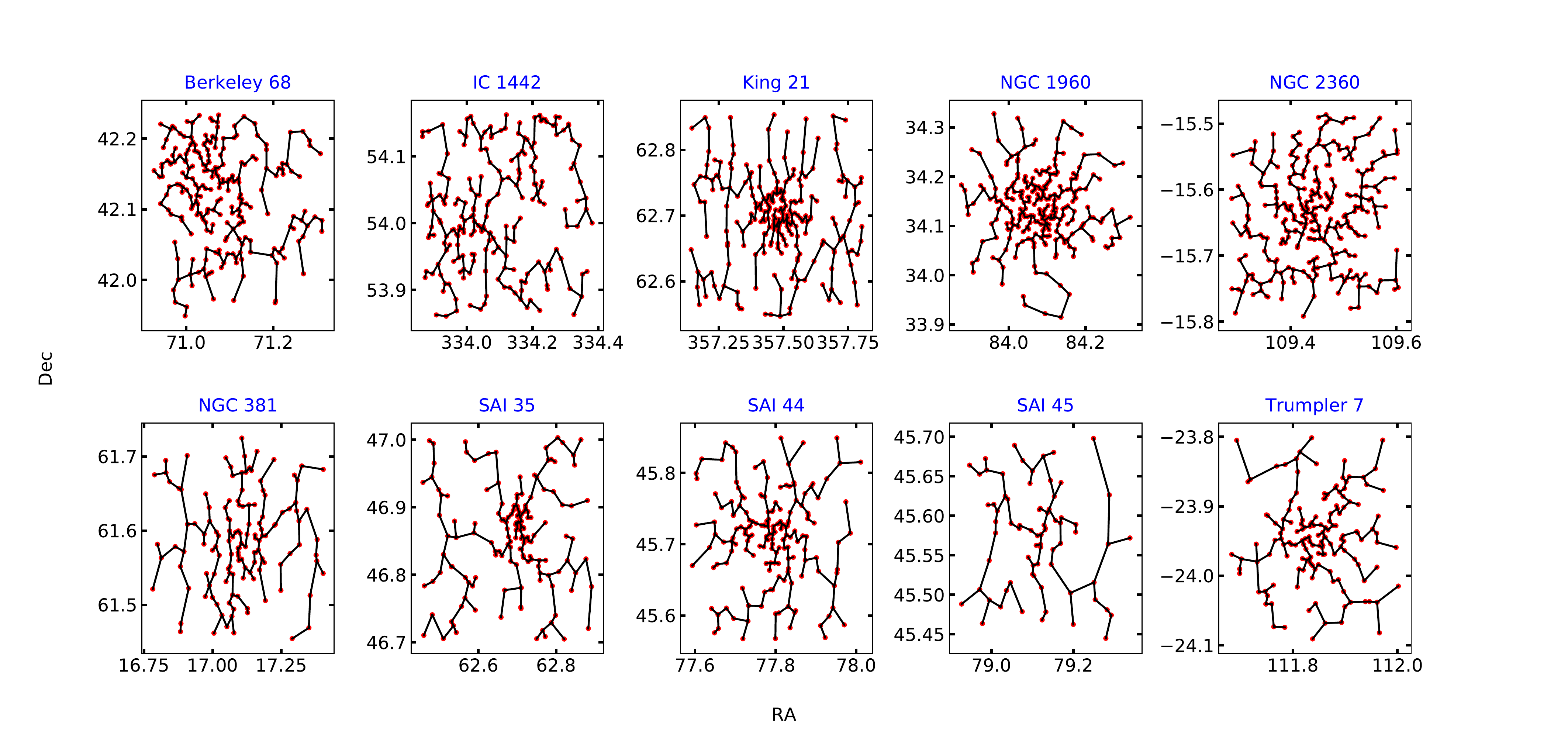}
\caption{Plot of minimum spanning tree for the clusters. The vertices (dots) in the plots represent member stars. The lines connecting these vertices are generally referred to as edges which are parts of a spanning tree}. The x and y axes in the plot represent the right ascension and declination for the stars associated with the clusters.
\label{mst}
\end{figure*}

The MSR is calculated from the mean edge length $\gamma$ using a minimum spanning tree (MST) for the member stars. The MST for the member stars is the shortest path that connects all the member stars barring closed loops \citep{1957BSTJ...36.1389P}. We have also shown the minimum spanning tree (MST) for these clusters in Figure~\ref{mst}. The MST was generated using the Python package provided by \citet{Naidoo2019}. The densely located stars would have smaller edge lengths in the MST. These MST plots are helpful in the visual inspection of the spatial distribution of the member stars in the clusters. The clusters King 21 and NGC 1960 show the concentration of vertices in their central region, however, this is related to the distribution of stars in general and cannot be directly linked to mass segregation as the MSTs shown  in Figure~\ref{mst} do not have stellar mass information.

The $\gamma$ is calculated for n most massive stars $\gamma_{mm}$ and n random stars chosen from all members of the cluster $\gamma_{rand}$, separately. Then we repeated the same process 500 times to calculate the mean of the $\gamma_{rand}$ i.e. $\langle \gamma_{rand} \rangle$. The MSR value, $\Gamma_{MSR}$, was calculated using the formula provided by \citet{2011A&A...532A.119O} which is given below:
$$
\Gamma_{MSR} = \frac{\langle \gamma_{rand} \rangle}{\gamma_{mm}}
$$
We calculated the standard deviation $\Delta\Gamma_{MSR}$ value in the mass segregation ratio $\Gamma_{MSR}$ as follows:
$$
\Delta\Gamma_{MSR} = \Delta\gamma_{rand}
$$

The MSR method works on the principle that mass segregation will cause massive stars to have closer spatial distribution than low-mass stars. The $\Gamma_{MSR}$ value for a cluster having similar spatial distribution for both the massive and low-mass stars will be $\sim$ 1 which means the absence of mass segregation. The value of $\Gamma_{MSR}$ greater than 1 would be interpreted as the relatively closer spatial distribution of the most massive stars than the rest of the stars which means the presence of mass segregation.

\begin{figure*}
\centering
\includegraphics[width=17 cm, height=15 cm]{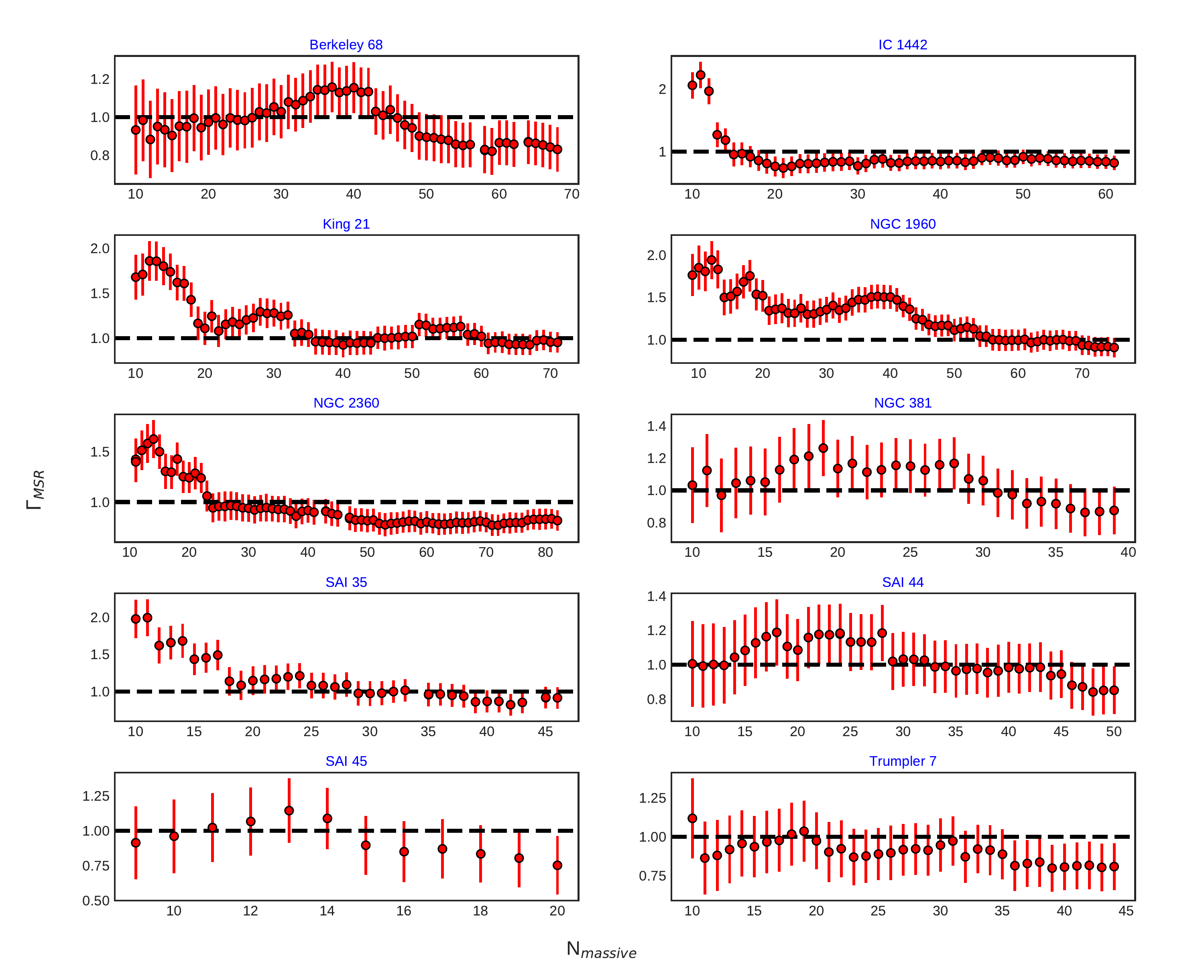}
\caption{Plot of $\Gamma_{MSR}$ vs N$_{massive}$ for the clusters. The error bar in the plot denotes the standard deviation $\Delta\Gamma_{MSR}$ values for each value of N$_{massive}$.}
\label{m_seg}
\end{figure*}

We used only those member stars of these clusters to study the mass segregation whose magnitudes are within the completeness limit in the V band as mentioned in Table~\ref{clust_par}. The clusters included in this study have ages greater than 20 Myr and are not embedded anymore in the star-forming regions so we took star magnitudes as the proxy for their masses. In our calculation of the $\Gamma_{MSR}$, we started with a minimum of 10 most massive stars in the clusters to have statistically significant calculations then we repeated the calculations of the $\Gamma_{MSR}$ by increasing the number of massive stars in steps of one in each iteration up to 30$\%$ of the total stars. The mass range, denoted by M$_{30 \%}$, corresponding to the most massive stars whose numbers are up to 30$\%$ of total stars is given in Table~\ref{rh_rt_t}.   We have shown the plots for the obtained values of $\Gamma_{MSR}$ versus the number of massive stars N$_{massive}$ in Figure~\ref{m_seg}.

We have reported the maximum value of $\Gamma_{MSR}$ obtained for the clusters in Table~\ref{rh_rt_t}. We found that the clusters NGC 2360, NGC 1960, IC 1442, King 21, and SAI 35 exhibit moderate signatures of mass segregation having $\Gamma_{MSR}$ to be 1.65$\pm$0.18, 1.94$\pm$0.22, 2.21$\pm$0.20, 1.84$\pm$0.23, and 1.96$\pm$0.25, respectively.  The $\Gamma_{MSR}$ $\sim$ 2.0 values for IC 1442 are only for N$_{massive}$ values of 11, 12, and 13. After these values, the mass segregation ratio drops sharply to around 1.0 which indicates that IC 1442 exhibits mass segregation for only a small number of the most massive stars. The value of $\Gamma_{MSR}$ is 1.26$\pm$0.18 for cluster NGC 381 which indicates the presence of weak mass segregation in the cluster. The remaining four clusters namely Berkeley 68, Trumpler 7, SAI 44, and SAI 45 exhibit  $\Gamma_{MSR}$ values within uncertainties to be 1, thus we do not find any evidence of mass segregation in these clusters. There is a bump around N$_{massive}$ values from 30 to 42 for the cluster Berkeley 68 which could be due to  the presence of subgroups of massive stars \citep{2017ApJ...840...91Y}. However, the mass segregation ratio for cluster Berkeley 68 is very low with   $\Gamma_{MSR}$ = 1.16$\pm$0.13. 

\subsection{Half mass radius, tidal radius, and cluster structure}
It has been found that tidal interactions can influence the structure and dynamical evolution of the open clusters \citep{2010MNRAS.402.1841C}. The ratio of half mass radius to the tidal radius, R$_{h}$, can be used as an indicator of the disruption of a cluster caused by tidal forces \citep{2021MNRAS.500.4338A}. The radial distance from the cluster center containing half of the total mass of the clusters is defined as half mass radius. The tidal radius, R$_{t}$, is defined as the radial distance from the cluster center where the tidal field of the Galaxy is balanced by the gravitational field of the cluster. We used the relation given by \citet{2000ApJ...545..301K} to calculate the tidal radius as follows:
$$
R_{t} = \left(\frac{M_{C}}{2 M_{G}}\right)^{1/3}\times R_{gc}
$$
The symbols M$_{C}$ and R$_{gc}$ in the above equation denote the total mass of the cluster and the distance to the cluster from the Galactic center, respectively. The mass of the Galaxy contained within the Galactocentric distance of the cluster is denoted by M$_{G}$ and calculated by using \citet{1987ARA&A..25..377G} relation given below:
$$
M_{G} = 2 \times 10^{8} M_{\odot} \left(\frac{R_{gc}}{30pc }\right)^{1.2}
$$
We used the above two equations to calculate the tidal radius of the clusters. The R$_{gc}$ values for the clusters were calculated using the relation R$_{gc}^{2}$=R$_{\odot}^{2}$+(d$\cos{\textbf{\textit{b}}}$)$^2$-2R$_{\odot}$d$\cos{\textbf{\textit{b}}}\cos{\textbf{\textit{l}}}$. The d and R$_{\odot}$ represent distances to clusters and the distance of the Sun from the Galactic center, respectively whereas \textbf{\textit{l}} and \textbf{\textit{b}} denote the Galactic longitude and latitude, respectively. We used R$_{\odot}$ as 8.2$\pm$0.10 kpc for the calculations as given by \citet{2019MNRAS.486.1167B}. The values of  R$_{t}$ and R$_{gc}$  are given in Table~\ref{rh_rt_t}. \\ 

The stellar initial mass function peaks at 0.5 M$_{\odot}$ whereas mass completeness limits are above 0.5  M$_{\odot}$ for all the clusters studied in this analysis. Therefore, we estimated the total stellar mass, M$_{C}$, of the clusters utilizing the stellar mass function. The mass function slopes for these clusters obtained in our previous studies for the stars above 1 M$_{\odot}$ were mostly in agreement with the \citet{2001MNRAS.322..231K} mass-function slopes within the uncertainty values \citep{2020MNRAS.492.3602J,2020MNRAS.494.4713M,2020MNRAS.495.2496M,2021AJ....162...64M}.  So, we used \citet{2001MNRAS.322..231K} mass function of multiple-part power law form to estimate the total mass of the clusters including stellar populations up to 0.08 M$_{\odot}$. We applied a similar method as described by \citet{2009ApJ...700..506S}. We are giving a brief description here.

The number of stars in the mass range m$_{1}$ to m$_{2}$ will be:

$$N = A\times \int_{m_{1}}^{m_{2}} M^{-\alpha} dM$$ 
Using the above equation for the known number of stars from our samples, we calculated the values of the normalization constant A for the $\alpha$ corresponding to the mass range of  M/M$_{\odot}$ $\geq$ 1 given by \citet{2001MNRAS.322..231K}.   The total mass can be obtained using the following relation:

$$M_{tot} = A\times \int_{0}^{N} M dN = A\times \int_{m_{1}}^{m_{2}} M^{1-\alpha} dM$$
Using the above equation, we calculated the total mass of the clusters in the mass ranges M/M$_{\odot}$ $\geq$ 1.0, 0.5 $\leq$ M/M$_{\odot}$ $<$ 1.0, 0.08 $\leq$ M/M$_{\odot}$ $<$ 0.5.  The normalization constant values for the other mass ranges namely 0.5 $\leq$ M/M$_{\odot}$ $<$ 1.0 and 0.08 $\leq$ M/M$_{\odot}$ $<$ 0.5 were calculated using the relations of the normalization constant values for different mass ranges given by \citet{2013MNRAS.429.1725M}. The estimated total mass values, M$_{C}$, through this method up to the lower stellar mass limit of 0.08 M$_{\odot}$  for the clusters are given in Table~\ref{rh_rt_t}.
\\

\begin{figure}
\centering
\includegraphics[width=9 cm, height=8.5 cm]{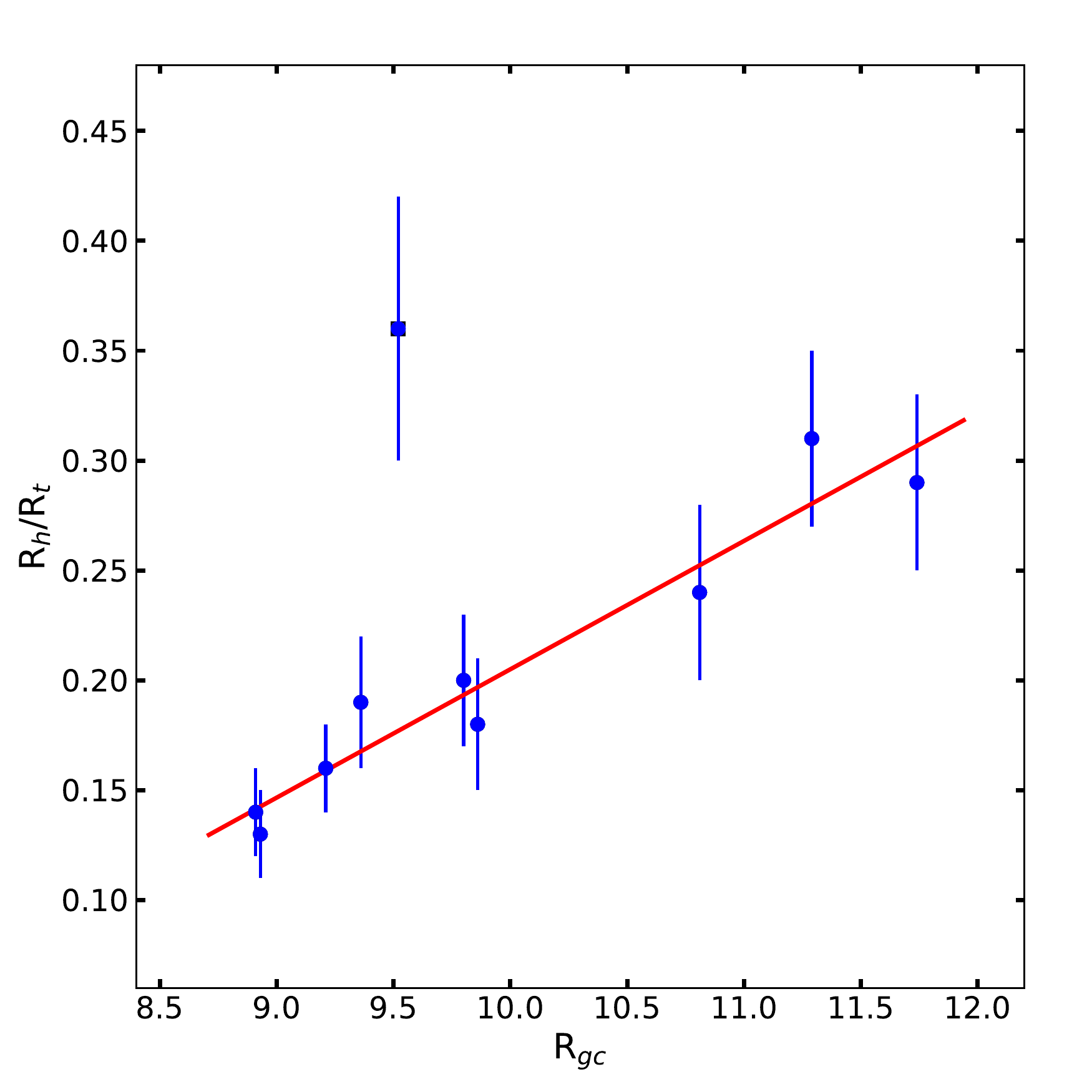}
\caption{Plot of R$_{h}$/R$_{t}$ vs R$_{gc}$ for the clusters. The point shown by the square is an outlier. The outlier point belongs to cluster IC 1442.}
\label{rh_rt_1}
\end{figure}

The half-mass radii, R$_{h}$, of the clusters were calculated from the individual masses of member stars of the clusters. The individual masses of the member stars were determined by fitting the solar metallicity isochrones of Marigo et al. (2017) on the V/(B-V) colour-magnitude diagrams of the clusters. These  isochrones were corresponding to the derived age, reddening, and distance of the clusters. The half-mass radii were estimated corresponding to the observed total masses of the clusters. We derived the half-mass radii of the clusters by including stars with magnitudes within the completeness limits of our data mentioned in Table~\ref{clust_par}.  Generally, the half-mass radii estimated from deep photometric data tend to become larger compared to the half-mass radii estimated from shallow photometric data \citep{2008A&A...477..829B}. However, half-mass radii and other structural parameters become insensitive to photometric depth in case of uniform mass function slopes i.e. absence of mass segregation \citep{2008A&A...477..829B}. As the studied clusters do not show strong mass segregation and the $\Gamma_{MSR}$ values reach around 1.0 even for most massive stars which are 30$\%$ of the total stars, the half-mass radii derived by us are reasonably good estimates. However, a larger sample having clusters with a high number of member stars would be better to constrain the correlation shown in Figure~\ref{rh_rt_1}.  The values of R$_{h}$ and R$_{h}$/R$_{t}$ are given in Table~\ref{rh_rt_t}.
\begin{table*}
  \centering
  \caption{The name of the clusters, the total mass of the clusters, the mass range of 30$\%$ most massive stars, R$_{h}$, R$_{t}$, R$_{gc}$, R$_{h}$/R$_{t}$, and $\Gamma_{MSR}$ are given in columns 1 to 8, respectively.}
  \label{rh_rt_t}
  \begin{tabular}{c c c c c c c c c c}  
  \hline  
    Cluster &  M$_{C}$& M$_{30 \%}$ & R$_{h}$& R$_{t}$& R$_{gc}$& R$_{h}$/R$_{t}$& $\Gamma_{MSR}$ \\
            & (M$_{\odot}$)& (M$_{\odot}$)&  (pc)   &  (pc)  &   (pc)   &    &   \\
  \hline
  NGC 381& 467.13$\pm$187.41& 2.80-1.86& 1.50$\pm$0.05& 10.30$\pm$1.36& 8.91$\pm$0.02& 0.15$\pm$0.02& 1.26$\pm$0.18 \\
  NGC 2360& 887.54$\pm$356.08& 2.21-1.17& 1.65$\pm$0.07& 12.75$\pm$1.69& 8.93$\pm$0.03& 0.13$\pm$0.02& 1.65$\pm$0.18 \\
  Berkeley 68    & 1396.11$\pm$552.29& 1.75-1.27& 5.41$\pm$0.34& 17.06$\pm$2.23& 11.29$\pm$0.19& 0.32$\pm$0.05& 1.16 $\pm$ 0.13 \\ 
  NGC 1960& 618.94$\pm$248.32& 7.18-1.51& 2.21$\pm$0.10& 11.65$\pm$1.54& 9.36$\pm$0.05& 0.19$\pm$0.03& 1.94$\pm$0.22 \\
  IC 1442& 1811.75$\pm$901.03& 9.02-2.38& 6.30$\pm$0.41& 16.77$\pm$2.76& 9.52$\pm$0.12& 0.38$\pm$0.07& 2.21$\pm$0.20 \\
  King 21& 1454.19$\pm$654.43& 6.89-4.22& 3.18$\pm$0.19& 15.93$\pm$2.37& 9.86$\pm$0.12& 0.20$\pm$0.03& 1.84$\pm$0.23 \\
  Trumpler 7& 508.00$\pm$203.81& 5.12-1.52& 1.78$\pm$0.10& 10.80$\pm$1.43& 9.21$\pm$0.06& 0.16$\pm$0.02& 1.04 $\pm$ 0.19 \\
  SAI 35& 817.47$\pm$327.96& 3.34-1.38& 3.78$\pm$0.36& 13.93$\pm$1.85& 10.81$\pm$0.25&  0.27$\pm$0.04& 1.96$\pm$0.25 \\
  SAI 44& 881.70$\pm$353.73& 2.47-1.39& 4.59$\pm$0.23& 15.01$\pm$1.99& 11.74$\pm$0.18& 0.31$\pm$0.04& 1.19$\pm$0.18 \\
  SAI 45& 356.18$\pm$142.90& 2.14-1.84& 2.09$\pm$0.06& 9.98$\pm$1.32& 9.80$\pm$0.05&  0.21$\pm$0.03& 1.14$\pm$0.24 \\
  \hline
  \end{tabular}
\end{table*}

We plotted R$_{h}$/R$_{t}$ versus R$_{gc}$ for the clusters as shown in Figure~\ref{rh_rt_1}. We found that the R$_{h}$/R$_{t}$ ratios are positively correlated with R$_{gc}$ values as can be seen in the figure. The R$_{h}$/R$_{t}$ value of 0.38$\pm$0.07 with R$_{gc}$ = 9.52$\pm$0.12 for the cluster IC 1442 is an outlier in the above plot. The binary clusters have been found to show relatively larger R$_{h}$/R$_{t}$ values \citep{2021MNRAS.500.4338A}.  The slope of the linear fit was found to be 0.06$\pm$0.01 with linear regression coefficient r-square = 0.93 in the range of R$_{gc}$ from 8.9 to 11.7 kpc. The clusters located at the larger Galactocentric distances might be subjected to weaker tidal fields which could be a possible reason for these clusters to have larger R$_{h}$/R$_{t}$ ratios \citep{2021MNRAS.500.4338A}.  

\section{Discussion}

We studied a sample of ten open clusters with photometric data mostly complete up to 19 mag in the V band to understand the dynamical evolution of the clusters. These clusters were chosen for being associated with comparatively lower reddening to assist the detection of fainter stars. We utilized precise astrometric data from Gaia DR3 to ascertain the membership of the stars and accurate measurements of the distances which helps in the precise estimation of distance modulus essential for mass determination through isochrone fitting on the CMDs of the clusters.

We used a method based on the minimum spanning tree of the member stars as suggested by \citet{2009MNRAS.395.1449A} to study mass segregation scenarios in the clusters. We find that the clusters NGC 2360, NGC 1960, IC 1442, King 21, and SAI 35 exhibit evidence of the presence of moderate mass segregation having mass segregation ratios to be 1.65$\pm$0.18, 1.94$\pm$0.22, 2.21$\pm$0.20, 1.84$\pm$0.23, and 1.96$\pm$0.25. The remaining clusters show weak or no mass segregation with $\Gamma_{MSR}$  around 1 within the uncertainty level. The mass segregation in the open clusters has been thought to increase with age \citep{2018MNRAS.473..849D}. In the present study, we could not find such a trend between mass segregation and the age of the clusters. Similarly, \citet{2022A&A...659A..59T} did not find any clear trend between the mass segregation ratios calculated from the ten most massive stars and the clusters' ages using a sample of 389 open clusters.

The ratio of the half-mass radius to the tidal radius, R$_{h}$/R$_{t}$, is a good parameter to study the influence of the tidal interactions on the dynamical evolution and shapes of the clusters \citep{2010MNRAS.401.1832B,2020MNRAS.493.3473A}. This ratio measures the fraction of the tidal volume filled by the half-mass contents. We estimated the total mass of the clusters for the calculation of  R$_{t}$ by using the multiple-part power law form of mass function given by \citet{2001MNRAS.322..231K} to include the masses of fainter stars up to 0.08 M$_{\odot}$. The ratios R$_{h}$/R$_{t}$ are found to be positively correlated to  R$_{gc}$ with a slope of 0.06$\pm$0.01 and linear regression r-square coefficient of 0.93. This indicates a tendency that the clusters located at larger R$_{gc}$ had larger  R$_{h}$/R$_{t}$ ratios. In a similar study with a larger sample of open clusters, \citet{2020MNRAS.493.3473A} could not find any such trend, however, they noticed that the clusters located at  R$_{gc}$ $>$ 9 kpc had at least 50$\%$ larger dispersion in R$_{h}$/R$_{t}$ values. It has been suggested that the clusters located at larger  R$_{gc}$ face lower external gravitational forces which allow the internal stellar content to fill a larger fraction of tidal volume without tidal disruption so these clusters exhibit larger R$_{h}$/R$_{t}$ values \citep{2020MNRAS.493.3473A,2021MNRAS.500.4338A}. The correlation between R$_{h}$/R$_{t}$ and R$_{gc}$, found here,  is based on a small sample of open clusters of a wide age range, which is an important caveat of the present study, therefore a study based  on larger samples of open clusters of similar age across different galactic locations would give more insight into  the dynamical evolution of the clusters.

\section{Conclusions}
We investigate the dynamical evolution of ten open clusters whose physical parameters were previously determined by us using a homogeneous approach and data sets. In the present study, we identified member stars of the clusters using Gaia DR3 astrometric data. Our findings can be summarized as follows: \\

\begin{itemize}
    \item We find moderate mass segregation with $\Gamma_{MSR}$ 
values to be 1.65$\pm$0.18, 1.94$\pm$0.22, 2.21$\pm$0.20, 1.84$\pm$0.23, and 1.96$\pm$0.25 for the clusters NGC 2360, NGC 1960, IC 1442, King 21, and SAI 35, respectively.
\item The cluster NGC 381 exhibits a weak signature of mass segregation with  $\Gamma_{MSR}$ is 1.26$\pm$0.18.
\item We find no evidence of mass segregation in the clusters Berkeley 68, Trumpler 7, SAI 44, and SAI 45.
\item The ratios R$_{h}$/R$_{t}$ are found to be positively correlated to  R$_{gc}$ with a slope of 0.06$\pm$0.01 and linear regression r-square coefficient of 0.93.
\end{itemize}


\balance


\bibliography{main.bib}{}
\bibliographystyle{aasjournal}

\end{document}